\documentclass[showpacs,amsmath,amssymb,onecolumn,pra,superscriptaddress]{revtex4-1}
\usepackage{amssymb}
\usepackage[dvips]{graphicx} % for figures
\usepackage{enumerate}
\usepackage{epsfig}
\usepackage{subfigure}
\usepackage{xcolor}
\usepackage[T1]{fontenc}
\usepackage[expert]{mathdesign}
\usepackage{fullpage}
\usepackage{amsthm,amsfonts,amssymb,amscd,mathrsfs,eufrak,xspace,framed}
\usepackage{amsmath}
\usepackage{color}
\usepackage{setspace}
\usepackage{url}
\usepackage{enumitem}
\usepackage{slashbox}

\newcommand{\bra}[1]{\mbox{$\left\langle #1 \right|$}}
\newcommand{\ket}[1]{\mbox{$\left| #1 \right\rangle$}}
\newcommand{\braket}[2]{\mbox{$\left\langle #1 | #2 \right\rangle$}}

\begin{document}

\title{Round-robin-differential-phase-shift quantum key distribution with and without monitoring signal disturbance}

\author{Zhen-Qiang Yin}
\author{Shuang Wang}
\email{wshuang@ustc.edu.cn}
\author{Wei Chen}
\email{weich@ustc.edu.cn}
\author{Yun-Guang Han}
\author{Rong Wang}
\author{Guang-Can Guo}
\author{Zheng-Fu Han}
\affiliation{CAS Key Laboratory of Quantum Information, University of Science and Technology of China, Hefei 230026, P. R. China}
\affiliation{Synergetic Innovation Center of Quantum Information $\&$ Quantum Physics, University of Science and Technology of China, Hefei, Anhui 230026, P. R. China}
\affiliation{State Key Laboratory of Cryptology, P. O. Box 5159, Beijing 100878, P. R. China}

\begin{abstract}
Round-robin-differential-phase (RRDPS) quantum key distribution (QKD) protocol has attracted intensive studies due to its distinct security characteristic, e.g., information leakage in RRDPS can be bounded without learning error rate of key bits. Nevertheless, its implementation is still far from practical due to the complication of its measurement device. Moreover, on the theoretical side, its security is still not clear in view of optimal attack. Here, by observing a potential phase randomization of the encoding states and its connection with eavesdropper's ancilla, we develop a theory to bound information leakage quite tightly and differently. Our theory is applicable for both with and without monitoring signal disturbance scenarios, which is significant for the understanding of RRDPS. Based on our novel security proof, the practicality and performance of RRDPS can be both improved dramatically. Furthermore, we realize a proof-of-principle experiment up to 140 km fiber distance which is the longest achievable distance of RRDPS system until now, while the original security proof predicts no secret key can be generated in our experiment. Our results pave an avenue towards practical RRDPS.
\end{abstract}

\maketitle

\newpage

\section{Introduction}
Unlike classical cryptography whose security relies on unproven mathematical assumptions, quantum key distribution (QKD) \cite{Bennett:BB84:1984,Ekert:QKD:1991} can information-theoretically distribute secret key bits between distant peers (such as Alice and Bob). According to quantum mechanics, any eavesdropping on quantum channel will inevitably introduce signal disturbance, which implies that Alice and Bob can bound the information leakage for the eavesdropper (Eve) through collecting the error rate of their raw key bits or some other parameters reflecting the signal disturbance. For the well known BB84 \cite{Bennett:BB84:1984} and Measurement-Device-Independent (MDI) \cite{Lo:MDIQKD:2012} QKD with decoy states \cite{Hwang:Decoy:2003,Wang:Decoy:2005,Lo:Decoy:2005}, the error rate and counting yields are used to evaluate Eve's information. In Coherent-One-Way (COW) \cite{cow,cow307} and Differential-Phase-Shift (DPS) \cite{DPS,DPS1} protocols, the visibility of interference plays the essential role to monitor information leakage. Device-Independent (DI) \cite{Acin:DeviceIn:07,Pironio:DeviceIn:09,Lim:DIQKD:2013} QKD also needs to observe the Bell inequality. MDI-QKD and DI-QKD feature their high security level in practice while COW and DPS have compact and simple implementation. There have been great progresses on experimental QKD, such as long distance distance QKD implementations \cite{cow307,MDI404,longdistance2017}, high key rate systems \cite{1M,DPS2,QKDWDM,MDI1G} and demonstrations of QKD network \cite{Peev:SECOQC:2009,Sasaki:TokyoQKD:2011,Frolich:QKDnet:2013,MDInetwork}.
Nevertheless, monitoring signal disturbance is indispensable for almost all these QKD protocols and implementations.

Surprisingly, recently proposed round-robin-differential-phase-shift (RRDPS) \cite{sasaki2014practical} protocol is an exception. In RRDPS protocol, Alice prepares a series of pulse trains, each consisting of $L$ weak coherent pulses. The pulses are individually modulated to random phases out of $0$ and $\pi$, and every $L$-pulse train can be handled as a packet. Upon receiving these packets, Bob measures the phase shift between the $i$-th pulse and $(i+r)$-th pulse of each packet, where $r$ is randomly chosen from $[1,L-1]$ for each packet and $i+r\leqslant L$. Through a simple and comprehensive security proof \cite{sasaki2014practical}, it has been pointed out that Eve's information on raw key bits $I_{AE}$ is no larger than $h_2(N/(L-1))$, where $N$ is the photon number of a packet. The main merit of the RRDPS protocol is that $I_{AE}$ does not depend on error rate of key bits, and thus can be treated as a constant experimentally. It's obvious that the information leakage will be deeply suppressed and higher tolerance of error rate is expected when $L$ becomes large, which is the reason why a RRDPS experiment with large $L$ is important. It is worth noting that multi-dimensional QKD protocols \cite{QKDwithd-level} usually have higher tolerance of error rate, especially the recently proposed Chau15 protocol \cite{chau15} can tolerate up to $50\%$ error rate in principle. However, these protocols must run with monitoring signal disturbance.

There have been several successful demonstrations of this protocol with passive interferometers \cite{RRDPSexp0,RRDPSexp1} and actively-selectable components \cite{RRDPSexp2,RRDPSexp3}. The longest achievable distance is around 90 km \cite{RRDPSexp3}.
Albeit great progresses on experiments of RRDPS protocol have been made, it's still a great challenge to realize a practical measurement system with large $L$ value. Besides, large $L$ value will decrease the secret key rate per pulse obviously. Therefore, it is highly desired if $I_{AE}$ can be further lowered while $L$ is maintained small. Additionally, although $I_{AE}$ given in Ref.\cite{sasaki2014practical} does not depend on the error rate, theorists are still not clear how does Eve's attack introduce error bits, and if it is possible to use the error rate in RRDPS to improve its performance. To address these issues, we first report a new theory to bound $I_{AE}$ greatly tighter than before especially for small $L$ values. Interestingly, error rate can be also taken into account in our method to estimate $I_{AE}$ further tightly. Through numerical simulation, we show that with our theory, the performance of the real-life RRDPS implementation can be improved dramatically. It is remarkable that the RRDPS protocol with $L=3$, which is not permitted in the original RRDPS protocol, can generate key bits now. Finally, we verify our theory through a proof-of-principle experiment with $L=3$, which achieves the longest achievable distance ($140$ km) so far.

\section{Results}
\subsection{New bound for Eve's information}
The original security proof given in Ref.\cite{sasaki2014practical} is simple and beautiful, but does not exploit Eve's optimal attack and corresponding information leakage. Our basic idea is to directly construct Eve's collective attack and calculate the maximal information acquired by Eve. Then using a quantum defitti theorem \cite{Caves:deFinetti:2002,definetti,Renner:deFinetti:2009}, the results are also against general coherent attacks. However, even for collective attack, it is not easy to analyze in the RRDPS protocol, since the dimension of Alice's encoding state depends on $L$ and may be very large. For simplicity, we first consider the case that each packet contains only one photon. Alice randomly prepares the single photon state $\ket{\psi}=\sum^{L}_{i=1}(-1)^{k_i}\ket{i}$, where $k_i\in\{0,1\}$ is Alice's raw key bit, and $\ket{i} (i\in\{1,..,L\})$ represents that this single photon is in the $i$-th time-bin. Eve's general collective attack can be given by $U_{Eve}\ket{i}\ket{e_{initial}}=\sum^L_{j=1}c_{ij}\ket{j}\ket{e_{ij}}$, where the quantum state of Eve's ancilla $\ket{e_{ij}}$ corresponds to that Eve transforms $\ket{i}$ to $\ket{j}$ and sends \ket{j} to Bob. In principle, Eve's ancilla has $L^2$ different states and thus is hard to tackle. We develop a method to simplify Eve's quantum state and calculate her information effectively. The essential of our method is to introduce the phase randomization, which is simply bypassed in previous works. Concretely speaking, consider the case that Bob has measured the incoming single photon with basis $\ket{a}\pm\ket{b}$ successfully, and announced $(a,b)$ publicly. Then Eve aims to guess $k_a+k_b$, it is evidently that for any $i\neq a,b$ the phase $(-1)^{k_i}$ is completely random to Eve, which implies that some mixed components $\big|c_{ia}\big|^2\ket{e_{ia}}\bra{e_{ia}}+\big|c_{ib}\big|^2\ket{e_{ib}}\bra{e_{ib}}$ ($i\neq a,b$) will emerge in the density matrix of Eve. These mixed components are definitely useless for Eve, thus can be bypassed and simplify the security proof notably. Accordingly, we find that $I_{AE}\leqslant max_{0\leqslant x \leqslant 1}\varphi((L-1)x,1-x)/(L-1)$, where $\varphi(x,y)=-x\log_2{x}-y\log_2{y}+(x+y)\log_2{(x+y)}$. Besides, $x$ can be related to the error rate $E$, thus this bound works for implementations both with and without monitoring signal disturbance, which is quite meaningful for practical QKD systems. One can refer to the supplementary file for a very detailed security proof.

It will be very useful to extend the single photon to $N$-photon case. Nevertheless, due to the complexity of $N$-photon quantum state, it is apparently very hard to depict and calculate Eve's information for general $N$-photon case. Our technique is to group the $N$-photon state into different summations with different number of phases and introduce phase randomization between them. Here, we sketch our method for the odd-$N$ photon-number case. Such an odd-$N$ photon quantum state must have the form $\ket{\psi}=\sum_{n=1}^{N/2+1/2}(-1)^{k_{i_1}+...+k_{i_{2n-1}}}\ket{i_1...i_{2n-1}}$, in which $\ket{i_1...i_{2n-1}}$ means a general state that the photon number in time-bins $i_1...i_{2n-1}$ must be odd, while the photon numbers in all other time-bins must be even. Then it is straightforward to redefine the collective attack with the new basis $\ket{i_1...i_{2n-1}}$: $U_{Eve}\ket{i_1...i_{2n-1}}\ket{e_{initial}}=\sum^L_{j=1}c_{i_1...i_{2n-1}j}\ket{j}\ket{e_{ii_1...i_{2n-1}j}}$. After Bob announces some $(a,b)$ publicly, Eve will try to guess $k_a+k_b$. Due to the potential phase randomization between different summations, Eve can acquire some information only from two types of "two-dimensional" terms like $U_{Eve}(-1)^{k_i}((-1)^{k_a}\ket{i_1a}+(-1)^{k_b}\ket{i_1b})\ket{e_{initial}}$ and $U_{Eve}(-1)^{k_i+k_j}(\ket{ij}+(-1)^{k_a+k_b}\ket{ijab})\ket{e_{initial}}$. Summing over Eve's information on all these "two-dimensional" terms, we obtain the final formula to estimate Eve's information. The detailed proof can be found in supplementary file. The results are summarized by the following theorem and its corollary.

{\it{Theorem.}} For the RRDPS protocol with $L$-pulse packet, each packet containing $N$ photon-number $(L\geqslant N+1)$, Eve's information can by bounded by
\begin{equation}
\label{Iaetext}
\begin{aligned}
I_{AE}\leqslant& Max_{x_1,x_2,...,x_{N+1}}\{\frac{\sum_{n=1}^{N}\varphi((L-n)x_n,nx_{n+1})}{L-1}\},
\end{aligned}
\end{equation}
where, $\varphi(x,y)=-x\log_2{x}-y\log_2{y}+(x+y)\log_2{(x+y)}$, non-negative real parameters $x_i$ satisfying $\sum_{i=1}^{N+1}x_i=1$. Moreover, if the error rate of raw key bits is $E$, these parameters $x_i$ must satisfy the constraint
\begin{equation}
\label{constext}
\begin{aligned}
&E\geqslant \frac{\sum^{(N-1)/2}_{n\geqslant 1}(\sqrt{(L-2n)x_{2n}}-\sqrt{2nx_{2n+1}})^2+(L-N-1)x_{N+1}/2}{L-1} \text{ for odd $N$,}\\
&E\geqslant \frac{\sum^{N/2}_{n\geqslant 1}(\sqrt{(L-2n+1)x_{2n-1}}-\sqrt{(2n-1)x_{2n}})^2+(L-N-1)x_{N+1}/2}{L-1} \text{ for even $N$}.\\
\end{aligned}
\end{equation}

{\it{Corollary.}} If the photon number $N\leqslant L-2$, $I_{AE}<1$ always holds.

 Based on this theorem, the estimation of $I_{AE}$ is generalized to find the maximum value of a given function under a constraint defined by $E$. This constraint can be simply bypassed, then we obtain $I_{AE}$ without monitoring signal disturbance. Alternatively, if we retain this constraint, a tighter estimation of $I_{AE}$ may be achieved. It is remarkable that searching such a maximum value can be effective and concise through numerical method, since this function is convex. We also note that in Ref.\cite{RRDPSma}, an improved estimation of $I_{AE}$ is also obtained. However, our bound is more tighter and does reveal the relation between information leakage and signal disturbance.

\subsection{Potential improvements made by our theory}

 We first compare the tolerance of $E$ between the original RRDPS and our new formulae. In Tab.\ref{table1}, under single photon case, the maximum tolerant error rates for RRDPS with conventional method and the proposed formulae are given. One can see that our formulae can increase the tolerance of error rate dramatically, especially when $L$ is small. It's remarkable to note that for the case $L=3$, our bound can tolerant $E$ up to $8\%$, while the original RRDPS protocol can not generate secure key bits at all. One may also note that the difference between these methods become little in large $L$ cases. The reason is quite simple, e.g., the original bound $h_2(1/(L-1))$ has been close to $0$ for large $L$, so the potential improvement made by our analyses will be very little.

\begin{table}[htb]
\centering
\caption{The maximum value of tolerant error rate of RRDPS with different methods.}
\label{table1}
\begin{tabular}{c|c|c|c}
\backslashbox {$L$}{method}  & original RRDPS & Eq.\eqref{Iaetext} without $E$ & Eq.\eqref{Iaetext} with $E$ \\
\hline
3 & -- & 0.0546 &  0.0811 \\
5 & 0.0289 & 0.122 & 0.144 \\
16 & 0.165 & 0.244 & 0.252 \\
32 & 0.24 & 0.3 & 0.303 \\
64 & 0.301 & 0.346 & 0.346 \\
\end{tabular}
\end{table}

  The most important thing is to compare the secret key rate and achievable distance between RRDPS protocols with the proposed $I_{AE}$ and original one. Our simulation is based on the realization given in Refs.\cite{RRDPSexp1,RRDPSexp2}. We simulate the secret key rates $R$ per pulse versus total losses for $L=16$, $L=32$ and $L=64$ without monitoring signal disturbance. In the simulation, we assume that dark counting rate $d=10^{-6}$ per pulse and the optical misalignment parameter $e_{mis}=0.015$ or $0.15$, which are typical and realistic. The detailed model used in simulation is given in the methods section.

\begin{figure}[h]
	\includegraphics[width=10cm]{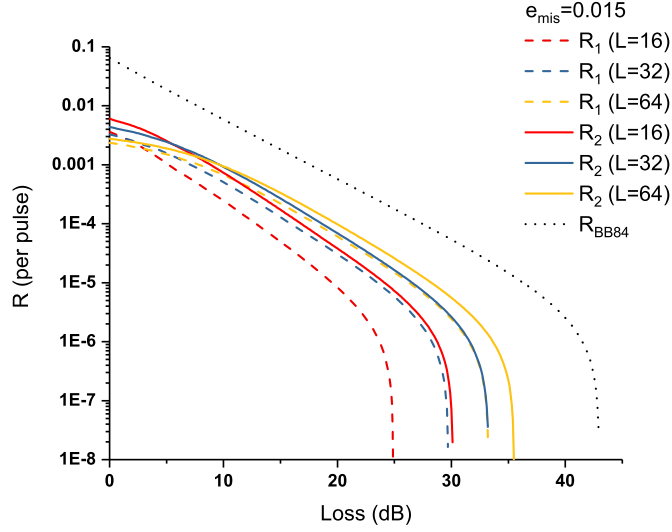}\\
\centering
	\caption{Secret key rate $R$ versus channel loss: $R_1$ and $R_2$ represent for the original RRDPS protocol and the proposed one respectively. $R_{BB84}$ is for the BB84 protocol with infinite decoy states. Both $R_1$ and $R_2$ are simulated for the scenarios without monitoring signal disturbance. }
	\label{comparison1}
\end{figure}

\begin{figure}[h]
	\includegraphics[width=10cm]{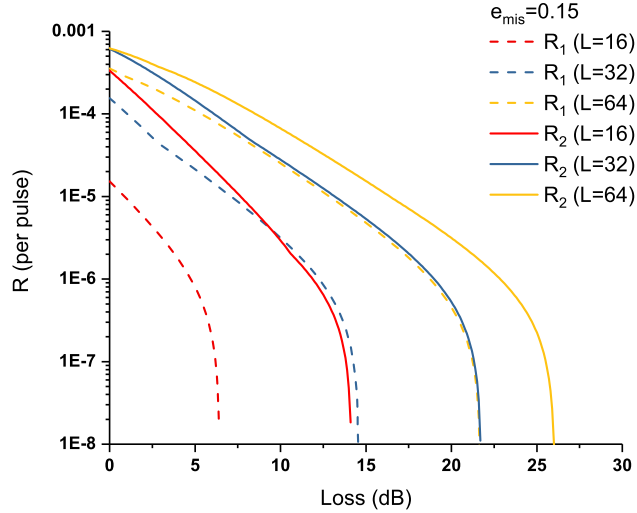}\\
\centering
	\caption{Secret key rate $R$ versus channel loss: $R_1$ and $R_2$ represent for the original RRDPS protocol and the proposed one respectively. $R_{BB84}$ is not drawn since $R_{BB84}=0$ in this case. Both $R_1$ and $R_2$ are simulated for the scenarios without monitoring signal disturbance.}
	\label{comparison2}
\end{figure}

Fig.\ref{comparison1} $(e_{mis}=0.015)$ and Fig.\ref{comparison2} $(e_{mis}=0.15)$ are both simulated without using signal disturbance parameters. From them we can see that with the help of the proposed method, the secret key rate and achievable distance of RRDPS systems are both evidently increased, especially for small $L$ cases. Although in low $e_{mis}$ case, the BB84 still overwhelms the proposed RRDPS in terms of key rate and achievable distance, the latter one runs without monitoring signal disturbance, which is very meaningful for the postprocessing of QKD systems. In high $e_{mis}$ example, the proposed RRDPS outperforms the BB84 significantly.

\begin{figure}[t]
	\includegraphics[width=10cm]{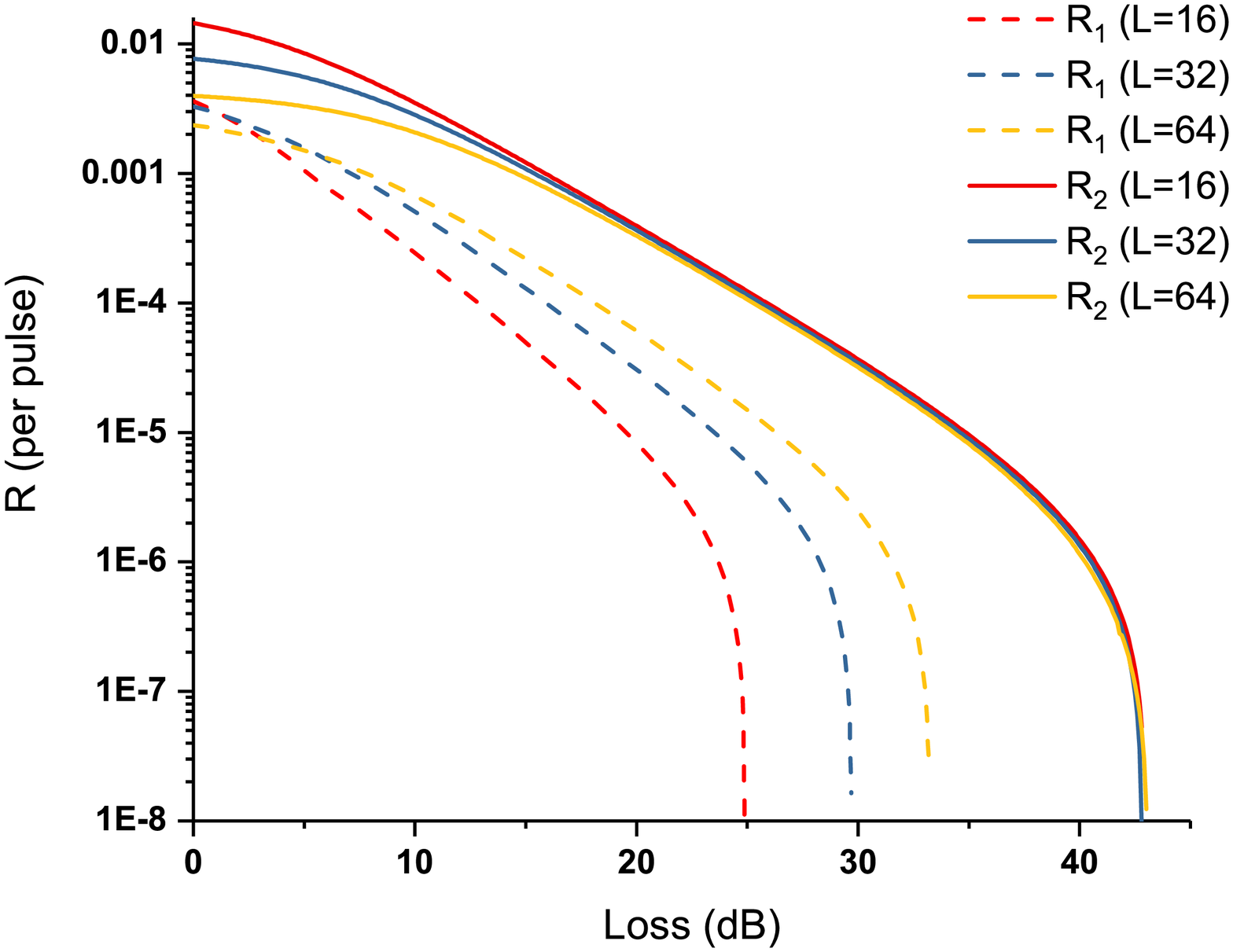}\\
	\caption{Secret key rate $R$ versus channel loss: $R_1$ represents for the original RRDPS protocol while $R_2$ means the proposed one with signal disturbance parameters.  $R_2$ is simulated with infinite decoy states.}
	\label{comparison3}
\end{figure}

We also simulate the RRDPS with monitoring signal disturbance in Fig.\ref{comparison3}, where we assume infinite decoy states \cite{Hwang:Decoy:2003,Wang:Decoy:2005,Lo:Decoy:2005} are employed and $e_{mis}=0.015$. The details of this simulation are given in the methods section.
With the error rate $E_i$ for the key bits generated from $i-$photon packet, the key rates are dramatically increased, especially for small $L$ cases. One may note that the achievable distance seems to be almost same for different $L$. Larger $L$ is, higher error rate from dark counts of single photon detectors will be introduced. Thus the achievable distance cannot increase unlimitedly through using larger $L$ in practice. Meanwhile, $I_{AE}$ is estimated very tightly, when the error rate is used. Based on the two points, the achievable distances for different $L$ in Fig.3 are very close.

Besides the numerical simulation, we borrow some experimental data to show the improvement of key rate. In an experiment of RRDPS with $L=65$ given in Ref.\cite{RRDPSexp2}, the secret key rate for $95km$ fiber channel can be increased from $5\times 10^{-8}$ to $1.4\times 10^{-6}$ per pulse (see methods section for details).

In conclusion, these simulations suggest that our theory can improve the performance of RRDPS protocol distinctly for scenarios both with and without monitoring signal disturbance. Especially, in the applications that optical interference is worse, i.e. high $e_{mis}$, our protocol overwhelms the commonly used BB84 protocol evidently.

\newpage

\subsection{Proof-of-principle experiment}
Based on the above theoretical results, the RRDPS requires that $L\geqslant 3$. From the view of experiment, $L=3$ corresponds to the most simple RRDPS realization. Here we make a proof-of-principle experiment with $L=3$ to verify our theory.

\begin{figure}[tbp]
	\includegraphics[width=14cm]{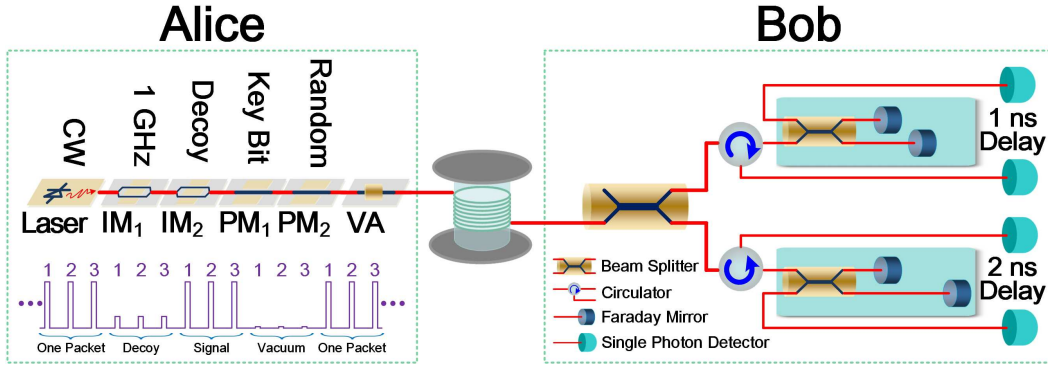}\\
	\caption{Experimental setup to implement RRDPS protocol with $L=3$. CW: Continuous Wave; IM: Intensity Modulator; PM: Phase Modulator; VA: Variable Attenuator.}
	\label{setup}
\end{figure}

Our implementation is shown in Fig.\ref{setup}, similar as \cite{RRDPSexp1,RRDPSexp2}. At Alice's site, a pulse train with the repetition rate of $1$ GHz is generated by modulating a $1550.12$ nm continuous wave (CW) laser using the first $LiNbO_3$ intensity modulator ($IM_1$). Every $3$ pulses ($L=3$) are defined as one packet. The second intensity modulator ($IM_2$) is employed to implement the decoy states method, by which each packet is randomly modulated into signal, decoy and vacuum packets. The first phase modulator ($PM_1$) adds phase $-\pi/2$ or $\pi/2$ on each pulse to encode the key bits, and the second phase modulator ($PM_2$) adds random global phase on each packet. The encoded pulse train is then launched into a variable attenuator (VA) so that the average photon number per pulse becomes the optimal value.

At Bob's site, the passive scheme based on a $1 \times 2$ beam splitter (BS) is used to implement a high-speed and low-loss decoding measurement. Since $L=3$ and the time interval between adjacent pulses is $1$ ns, there are only two unbalanced Faraday-Michelson interferometer (FMI) with $1$ ns and $2$ ns temporal delays. One $50/50$ BS and two Faraday mirrors (FM) constitute a FMI, and a three-port optical circulator is added before the BS to export the other interference result. Each output of these two unbalanced FMIs is led to a SPD. Finally, the detection events are recorded by a time-to-digital convertor (TDC), which records the time-tagged and which-detector information.

The passive implementation scheme and small value of $L$ and t make our RRDPS system very practical. The passive approach can achieves high time efficiency and internal transmittance of Bob's optical components, and four SPDs used to detect the $L=3$ packet is acceptable. The $1 \times 2$ BS amounts to randomly choosing between $1$ ns and $2$ ns delay FMIs. Different from active schemes \cite{RRDPSexp2}, the passive choice between different delay measurements has no speed limits, and the time interval between each two packets to achieve low error rate is not essential any more. The average insertion loss (IL) of the $1$ ns and $2$ ns delay FMIs is only approximate $0.80$ dB, where the IL of the optical circulator is also included. These two FMIs are placed in two small ABS plastic cases to isolate them from the environment, and heating plates are used to keep the temperature of FMIs above the room temperature. Thus, we could actively and independently compensate the phase shifts of $1$ ns and $2$ ns delay FMIs, and also keep the phase of the unbalance interferometer stable. Owing to $45^{\circ}$ Faraday mirrors, these FMIs is insensitive to polarization variations, and features extinction ratios of approximate $23.5$ dB.

In addition, SPDs based on InGaAs/InP avalanche photodiodes (APD) are employed to detect photons from $1$ ns and $2$ ns delay FMIs, which makes the RRDPS system more practical. These four SPDs are working with Peltier cooling, and operated in gated Geiger mode with the sine-wave filtering method \cite{sinspd}. The detection efficiencies of the four SPDs are approximately $20.4\%$ with a dark count rate of $1.25 \times 10^{-6}$ per gate and an after-pulse probability of $1.02\%$. Here, the insertion loss of the optical circular from BS (of FMI) to SPD is included in the detection efficiency of SPD.

We tested the $L=3$ RRDPS system with standard telecom fiber channels at the distance of $50$ km, $100$ km, and $140$ km. The decoy states method was implemented by setting the photons per pulse of signal, decoy, and "vacuum" packets with the value of 0.13, 0.03, and 0.0003, respectively. The experimental results are listed in Tab.II, where the error rates of key bits and yields per packet are directly obtained experimentally. Then we use formulae given in \cite{MXF:Practical:2005} to calculate the yield and error rate for single photon packet. Finally the secret key rates $R_1$ and $R_2$ are calculated according to Eq.\eqref{Iaetext} without and with error rate respectively (see methods section for detailed information).

For our $L=3$ RRDPS experiment system, the transmission distance could reach $140$ km with InGaAs/InP SPDs, while the maximum transmission distance of the similar $L=5$ RRDPS experiment system is less than $50$ km with superconducting SPDs \cite{RRDPSexp1}. Thus, we have successfully verified the feasibility of RRDPS with the smallest $L=3$, which is impossible based on original theory. However, this proof-of-principle experiment does not provide advantage over the commonly used QKD protocols, e.g. BB84 with decoy states, Indeed, to obtain distance or rate advantage over BB84 and bypass decoy states, a larger $L$ is necessary.

\begin{table}[!h]\center
\caption {Experimental results of the $L=3$ RRDPS system. List of mean yields and error rates of signal ($Q_s$ and $E_s$), decoy ($Q_d$ and $E_d$), and "vacuum" ($Q_v$ and $E_v$) packets, secure key rates per pulse ($R_1$ and $R_{2}$) for three lengths of the fiber channel ($l$), where $R_1$ is calculated without using error rate while $R_2$ is based on monitoring error rate.} \label{results}
	
\begin{tabular}{c c c c c c c c }
\hline $l$(km) & $Q_s$               & $E_s$    & $Q_d$               & $E_d$  &$Q_v$               & $R_{1}$          & $R_{2}$          \\
\hline $50$    & $3.24\times10^{-3}$ & $1.76\%$ & $7.52\times10^{-4}$ & 1.95\% &$1.12\times10^{-5}$  & $8.14\times10^{-5}$ & $3.60\times10^{-4}$ \\
\hline $100$   & $3.28\times10^{-4}$ & $2.26\%$ & $7.86\times10^{-5}$ & 4.01\% &$4.50\times10^{-6}$  & $4.98\times10^{-6}$ & $3.15\times10^{-5}$  \\
\hline $140$   & $5.52\times10^{-5}$ & $4.99\%$ & $1.56\times10^{-5}$ & 13.31\%&$3.87\times10^{-6}$  &  --               & $1.45\times10^{-6}$  \\
\hline
\end{tabular}

\end{table}

\section{Discussion}
We develop a theory to estimate Eve's information on raw key bits $I_{AE}$ in a quite different way. Briefly speaking, the new physics behind our method is that the potential phase randomization can be utilized for the security analysis of RRDPS. The main merit of our method is that $I_{AE}$ could be bounded more tightly than before, especially when $L$ is small. In theory, the relation between the information leakage and error rate in RRDPS is present clearly, which is particularly meaningful for the completeness of security analysis of QKD. Our results can be used for scenarios both with and without monitoring signal disturbance. If the error rate $E$ is known, a more precise estimation of Eve's information can be established. With the help of our theory, the secret key rate and achievable distance of RRDPS are both improved greatly for implementations both with and without monitoring signal disturbance. We also compare the RRDPS with the commonly used BB84 protocol. In terms of secret key rate and achievable distance, the RRDPS may not be better than BB84 when optical misalignment is low. Nevertheless, it can be run without monitoring signal disturbance, thus to be useful in many applications. Moreover, when optical misalignment is high, e.g. poor active phase compensation due to disturbance, RRDPS can outperform the BB84 significantly. To verify our theory, a proof-of-principle experiment with $L=3$ is demonstrated here.

Nevertheless, there are still several issues may be addressed in future. In Ref.\cite{RRDPSwithsourceflaws}, it has been proved that the original bound $I_{AE}\leqslant h_2(n/(L-1))$ holds with inaccurate phase coding. However, our technique deeply depends on the phase randomization which requires that Alice's phase coding must be $0$ or $\pi$ randomly. Therefore, analyzing the relation between phase coding inaccuracy and $I_{AE}$ quantitatively is necessary. Another issue is how to countermeasure the potential attacks due to device imperfections. For example, the blinding attack \cite{Lydersen:Hacking:2010} must be carefully considered in the practical RRDPS systems.

Note added. While preparing the paper, we are aware that similar topics are discussed in theoretical works \cite{RRDPSwithsmalldelays,RRDPSwitherrorrate}. The methods used here are completely different from theirs. Compared to Ref.\cite{RRDPSwithsmalldelays}, our theory can effectively estimate $I_{AE}$ without monitoring signal disturbance. As for \cite{RRDPSwitherrorrate}, our results obtain a more tighter bound of $I_{AE}$.

\section{Methods}
\subsection{Simulation}
We use the Wolfram Mathematica 10.3 to run the numerical simulations. The models of the simulations are given bellow.

{\it{Without monitoring signal disturbance.}} Assuming the mean photon number
of each pulse prepared by Alice to be $\mu$, its mean photon number becomes $\eta\mu$ after traveling through the channel with transmission efficiency $\eta$.
When Bob decides to set the delay value as $r\in\{1,...,L-1\}$, his photon number resolving SPDs will open $L-r$ time-windows to detect the incoming signal. Bob only retains the events that just one single photon click occurs among these $L-r$ time-windows. Here we assume that all loss stems from the channel while the photon-number resolving SPD have $100\%$ efficiency and dark counting rate $d$ per pulse. Accordingly, we obtain the counting rate with delay value $r$
\begin{equation}
\begin{aligned}
Q_r=(1-d)^{2(L-r)-1}e^{-(L-r)\eta\mu}((L-r)\eta\mu+2(L-r)d),
\end{aligned}
\end{equation}
and the overall counting rate $Q=\sum^{L-1}_{r=1}Q_r/(L-1)$. The error rate $E$ can be simulated by
\begin{equation}
EQ=\sum^{L-1}_{r=1}\frac{1}{L-1}(1-d)^{2(L-r)-1}e^{-(L-r)\eta\mu}((L-r)\eta\mu e_{mis}+(L-r)d).
\end{equation}
In the case without monitoring signal disturbance and decoy states, the secret key rate $R$ per pulse is given by
\begin{equation}
\begin{aligned}
\label{Rwcs}
RL=Q(1-h_2(E))-e_{src}-(Q-e_{src})I_{AE},
\end{aligned}
\end{equation}
where, $e_{src}=1-\sum_{i=0}^{v_{th}}e^{-L \mu}(L\mu)^i/i!$ is the probability of the photon number of a packet is larger than $\nu_{th}$, $h_2$ is the information entropy function. In our method $I_{AE}$ is calculated by Eq.\eqref{Iaetext} setting $N=v_{th}$ and ignoring constraint Eq.\eqref{constext}. In the original method, $I_{AE}=h_2(v_{th}/(L-1))$. $\mu$ and $v_{th}$ should be optimized to achieve the maximum $R$.

{\it{With monitoring signal disturbance.}} When signal-disturbance-monitoring and infinite decoy states are both active,
\begin{equation}
\begin{aligned}
\label{Rwcs}
RL=Q(1-h_2(E))-\sum_{i=1}e^{-L \mu}\frac{(L\mu)^i}{i!} Y_iI_{AEi},
\end{aligned}
\end{equation}
where $Y_i$ is the yield for $i$-photon trial, $I_{AEi}$ is Eve's information for the key bit generated by $i$-photon trial. In our method $I_{AEi}$ is calculated by Eq.\eqref{Iaetext} setting $N=i$ and constraint given by Eq.\eqref{constext}. Note that we set $I_{AEi}=1$ for $i\geqslant L-1$.

In this case, we first simulate the probability that Bob obtains one raw key bit per $i$-photon packet with delay value $r$, which is
\begin{equation}
\begin{aligned}
Y_{ir}&=(1-\frac{L-r}{L}\eta)^{i-1}(1-d)^{2(L-r)-1}(\frac{L-r}{L}i\eta+(1-\frac{L-r}{L}\eta)2(L-r)d).\\
\end{aligned}
\end{equation}
And the error rate of key bit generated by $i$-photon packet with delay value $r$ is

\begin{equation}
\begin{aligned}
E_{ir}Y_{ir}&=(1-\frac{L-r}{L}\eta)^{i-1}(1-d)^{2(L-r)-1}(\frac{L-r}{L}i\eta e_{mis}+(1-\frac{L-r}{L}\eta)(L-r)d),\\
\end{aligned}
\end{equation}
where, $e_{mis}$ represents the probability that the incoming photon clicks the erroneous SPD due to optical misalignment.
Accordingly, the mean yield of an $i$-photon packet is $Y_i=\sum^{L-1}_{r=1}Y_{ir}/(L-1)$, its mean error rate is simulated by $E_iY_i=\sum^{L-1}_{r=1}E_{ir}Y_{ir}/(L-1)$. Now we are ready to simulate RRDPS with monitoring signal disturbance.

{\it{Simulation of BB84.}} For the purpose of comparison, we also simulate phase-coding BB84 with infinite decoy states here. For BB84,
\begin{equation}
\begin{aligned}
&Y_{i}=(1-\frac{1}{2}\eta)^{i-1}(1-d)(\frac{1}{2}i\eta+(1-\frac{1}{2}\eta)2d),\\
&E_iY_{i}=(1-\frac{1}{2}\eta)^{i-1}(1-d)(\frac{1}{2}i\eta e_{mis}+(1-\frac{1}{2}\eta)d).
\end{aligned}
\end{equation}
And its key rate is
\begin{equation}
\begin{aligned}
\label{RBB84wcs}
2 R=-Qh_2(E)+e^{-2\mu}2\mu Y_1(1-h_2(E_1)).
\end{aligned}
\end{equation}

{\it{Calculations for an existing experiment.}} In an experiment of RRDPS with $L=65$ given in Ref.\cite{RRDPSexp2}, there is a set of experimental observations: the mean photon number $s=0.037$ per pulse, the yield $Q_s=8.435\times 10^{-4}$ per packet and error rate $E=0.058$. By setting $v_{th}=10$, the secret key rate is $R_1=(Q_s(1-1.1h_2(E))-e_{src}-(Q_s-e_{src})h_2(v_{th}/64))/L=5\times 10^{-8}$. With the same parameters and finding $I_{AE}=0.513$ for 10-photon, $R_2=(Q_s(1-1.1h_2(E))-e_{src}-(Q_s-e_{src})I_{AE})/L=1.44\times 10^{-6}$.

\subsection{Key rate for the experiment}
Here we give the methods how to get the secret key rates in Tab.II. The photons per pulse of signal , decoy, and "vacuum" packets are setting with the values of $s=0.13$, $d=0.03$, and $v=0.0003$ respectively. In experiment, we directly observe the yields $Q_s$, $Q_{d}$ and $Q_{v}$ for signal, decoy and "vacuum" packets respectively. The error rate $E_s$ $(E_d)$ for key bits generated from signal (decoy) packets are also observed experimentally.  Refer to Ref.\cite{MXF:Practical:2005}, we can estimate the yield $Y_1$ for packets with single photon and the error rate $E_1$ for key bits generated from single photon packets by the followings:
\begin{equation}
\begin{aligned}
\label{Y1E1}
&Y_0=max\{\frac{LdQ_v e^{L v} - L v Q_d e^{L d}}{L d - L v},0\}£¬\\
&Y_1=\frac{Ls}{LsLd-LsLv-(Ld)^2+(Lv)^2}(Q_de^d-Q_ve^v-\frac{(Ld)^2-(Lv)^2}{(Ls)^2}(Q_se^s-Y_0)),\\
&E_1=\frac{E_s Q_s e^{L s} - E_d Q_d e^{L d}}{(Ls-Ld)Y_1}.
\end{aligned}
\end{equation}
As a proof-of-principle experiment, the secret key rates $R_1$ and $R_2$ in Tab.II are not obtained by actually performing post-processing steps. Instead, they are calculated by $R=(Lse^{-Ls}Y_1 (1-I_{AE})-Q_sh_2(E_s))/L$. Here, to obtain $R_1$ we calculate $I_{AE}$ with Eq.\eqref{Iaetext} ignoring constraint Eq.\eqref{constext}. To $R_2$, this constraint with $E=E_1$ is used.

\bibliographystyle{apsrev4-1}

\section{Acknowledgements}
The authors thank Prof. Xiongfeng Ma, Dr. Xiao Yuan and Dr. Zhu Cao for helpful discussions. This work has been supported by the National Natural Science Foundation of China (Grant Nos. 61475148, 61627820, 61622506, 61575183, 61675189), the National Key Research And Development Program of China (Grant Nos.2016YFA0302600, 2016YFA0301702), the "Strategic  Priority Research Program(B)" of the Chinese Academy of Sciences (Grant No. XDB01030100).

\bibliography{Biblisource}
\newpage

\section{Supplementary File: Detailed Security Proof}

\subsection{Single photon case}
Alice randomly prepares the single photon state $\ket{\psi}=\sum^{L}_{i=1}(-1)^{k_i}\ket{i}$, where $k_i\in\{0,1\}$ is Alice's raw key bit, and $\ket{i} (i\in\{1,..,L\})$ represents that a single photon is in the $i$-th time-bin. Eve's general collective attack can be given by:

\begin{equation}
\label{attack}
\begin{aligned}
&U_{Eve}\ket{i}\ket{e_{00}}=\sum^L_{j=1}c_{ij}\ket{j}\ket{e_{ij}}\\
\end{aligned}
\end{equation}
where, $\ket{e_{ij}}$ is the quantum state of Eve's ancilla. Without loss of generality, we assume $c_{ij}\geqslant 0$ and $\sum^L_{j=1}c_{ij}^2\leqslant 1$, where the reason of setting $\sum^L_{j=1}c_{ij}^2\leqslant 1$ is that Eve may introduce vacuum state. For each trial, Eve only retains her ancilla $\ket{e_{ij}}$ to obtain maximum information on key bits.

In RRDPS protocol, Bob measures the phase shift between $\ket{i}$ and $\ket{j}$ of the incoming single photon states. If Bob projects the incoming single photon states into $(\ket{a}\pm\ket{b})/\sqrt{2}$ successfully, he will announce $\{a,b\}(a<b)$ to Alice, who will calculate $k_a\oplus k_b$ as her sifted key. The evolution of quantum state will be

\begin{equation}
\label{evolution1}
\begin{aligned}
\ket{\psi}\ket{e_{00}}\longrightarrow &(-1)^{k_a}(\tilde{c}_{aa}\ket{a}+\tilde{c}_{ab}\ket{b})+(-1)^{k_b}(\tilde{c}_{bb}\ket{b}+\tilde{c}_{ba}\ket{a})\\
&+\sum_{i\neq a,b}(-1)^{k_i}(\tilde{c}_{ia}\ket{a}+\tilde{c}_{ib}\ket{b}),\\
\end{aligned}
\end{equation}
where, $\tilde{c}_{ij}\triangleq c_{ij}\ket{e_{ij}}$.

The density matrix (non-normallized) of Eve's ancilla will be

\begin{equation}
\label{rho}
\begin{aligned}
\rho_E&=P\{\frac{1}{\sqrt{2}}(\bra{a}+\bra{b})U_{Eve}\ket{\phi}\ket{e_{00}}\}+P\{\frac{1}{\sqrt{2}}(\bra{a}-\bra{b})U_{Eve}\ket{\phi}\ket{e_{00}}\}\\
&=P\{\sum^L_{i=1}(-1)^{k_i}\tilde{c}_{ia}\}+P\{\sum^L_{i=1}(-1)^{k_i}\tilde{c}_{ib}\},\\
\end{aligned}
\end{equation}
where $P\{\ket{x}\}=\ket{x}\bra{x}$.
Eve aims to guess $k_a\oplus k_b$ after Bob reveals the values of $a$ and $b$.

Next, we try to simplify Eve's density matrix. Since $k_i(i\neq a,b)$ equals to $0,1$ randomly, the relative phase between $\ket{e_{aa}}(\ket{e_{bb}})$ and $\ket{e_{ia}}(\ket{e_{bb}}),i\neq a,b$, will be randomized. In other words, we have the following consideration

\begin{equation}
\label{psia}
\begin{aligned}
\rho_E&\longrightarrow\sum_{j\neq a,b}\sum_{k_{j}=0,1}\rho_E\\
&=P\{(-1)^{k_a}\tilde{c}_{aa}+(-1)^{k_b}\tilde{c}_{ba}\}+P\{(-1)^{k_b}\tilde{c}_{bb}+(-1)^{k_a}\tilde{c}_{ab}\}\\
&+\sum_{i\neq a,b}c^2_{ia}P\{\ket{e_{ia}}\}+c^2_{ib}P\{\ket{e_{ib}}\}.\\
\end{aligned}
\end{equation}

Based on the above equation, if $k_a+k_b=0$, the density matrix (non-normalized) of Eve's ancilla $\ket{e}$ will be
\begin{equation}
\label{rho0}
\begin{aligned}
\rho^{(a,b)}_0&=P\{\tilde{c}_{aa}+\tilde{c}_{ba}\}+P\{\tilde{c}_{bb}+\tilde{c}_{ab}\}+\sum_{i\neq a,b}c^2_{ia}P\{\ket{e_{ia}}\}+c^2_{ib}P\{\ket{e_{ib}}\}.\\
\end{aligned}
\end{equation}
If $k_a+k_b=1$, the density matrix of Eve's ancilla $\ket{e}$ will be
\begin{equation}
\label{rho1}
\begin{aligned}
\rho^{(a,b)}_1&=P\{\tilde{c}_{aa}-\tilde{c}_{ba}\}+P\{\tilde{c}_{bb}-\tilde{c}_{ab}\}+\sum_{i\neq a,b}c^2_{ia}P\{\ket{e_{ia}}\}+c^2_{ib}P\{\ket{e_{ib}}\}.\\
\end{aligned}
\end{equation}

Without compromising the security, we can assume that $\braket{e_{im}}{e_{jn}}=\delta_{ij}\delta_{mn}$.
Then, Eve's information on $k_a\oplus k_b$ is given by the Holevo bound \cite{Holevo:Bound:1973}, which is
\begin{equation}
\label{Iab}
\begin{aligned}
Q^{(a,b)}I^{(a,b)}_{AE}&=(c^2_{aa}+c^2_{ba})S(\left[
\begin{matrix}
\frac{c^2_{aa}}{c^2_{aa}+c^2_{ba}}&0\\
0&\frac{c^2_{ba}}{c^2_{aa}+c^2_{ba}}\\
\end{matrix}
\right])+(c^2_{bb}+c^2_{ab})S(\left[
\begin{matrix}
\frac{c^2_{bb}}{c^2_{bb}+c^2_{ab}}&0\\
0&\frac{c^2_{ab}}{c^2_{bb}+c^2_{ab}}\\
\end{matrix}
\right])\\
&=\varphi(c^2_{ba}, c^2_{aa})+\varphi(c^2_{ab},c^2_{bb}),\\
\end{aligned}
\end{equation}
where, $Q^{(a,b)}=\sum_{i}(c^2_{ia}+c^2_{ib})$ is the yield for any $a,b$, $\varphi(x^2,y^2)=-x^2\log_2{x^2}-y^2\log_2{y^2}+(x^2+y^2)\log_2{(x^2+y^2)}$.
Thus Eve's information on raw key bit is

\begin{equation}
\label{I1}
\begin{aligned}
I_{AE}&=\frac{\sum_{a<b}Q^{(a,b)}I^{(a,b)}_{AE}}{\sum_{a<b}Q^{(a,b)}}=\frac{\sum_{a<b}\varphi(c^2_{ba}, c^2_{aa})+\varphi(c^2_{ab},c^2_{bb})}{(L-1){\sum_{i,j}c^2_{ij}}}.
\end{aligned}
\end{equation}
 Note that $\varphi(x^2,y^2)$ is a concave function, then using the Jensen's inequality, we have

\begin{equation}
\begin{aligned}
\sum_{a<b}\varphi(c^2_{ba}, c^2_{aa})+\varphi(c^2_{ab},c^2_{bb})& \leq \varphi(\sum_{a<b}c^2_{aa}+c^2_{bb},\sum_{a<b}c^2_{ba}+c^2_{ab}))=\varphi((L-1)\sum_{i}c^2_{ii},\sum_{i\neq j}c^2_{ij})\\
&= \varphi((L-1)x_1,x_2),\\
\end{aligned}
\end{equation}
where, we define $x_1=\sum_{i}c^2_{ii}$ and $x_2=\sum_{i\neq j}c^2_{ij}$.
Consequently, we have
\begin{equation}
\label{Iae1}
\begin{aligned}
I_{AE}&\leqslant \frac{\varphi((L-1)x_1,x_2)}{(L-1)(x_1+x_2)}.
\end{aligned}
\end{equation}
By searching the maximum of above function with free non-negative variables $x_1$ and $x_2$ ($x_1+x_2>0$), we can obtain the maximal information leaked to Eve. Next, we try to bound $I_{AE}$ further tightly by finding the relationship between $x_1$ and $x_2$.

Intuitively, the parameters $x_1$ and $x_2$ may depend on the error rate of the sifted key bit. In the following, we try to introduce the error rate into the security proof of RRDPS protocol. According to Eq.\eqref{evolution1}, when $k_a+k_b=0$, the probability of Bob obtaining an error bit is
\begin{equation}
\begin{aligned}
p^{(a,b)}_e=\frac{1}{2}[\big| \tilde{c}_{aa}+\tilde{c}_{ba}-\tilde{c}_{ab}-\tilde{c}_{bb}\big|^2+\sum_{i\neq a,b}\big|\tilde{c}_{ia}-\tilde{c}_{ib}\big|^2].
\end{aligned}
\end{equation}

 In the case that $k_a + k_b=1$, the probability of Bob obtaining $(\ket{a}+\ket{b})/\sqrt{2}$ is
\begin{equation}
\begin{aligned}
p'^{(a,b)}_e=\frac{1}{2}[\big| \tilde{c}_{aa}-\tilde{c}_{ba}+\tilde{c}_{ab}-\tilde{c}_{bb}\big|^2+\sum_{i\neq a,b}\big|\tilde{c}_{ia}+\tilde{c}_{ib}\big|^2].
\end{aligned}
\end{equation}

We are ready to give the relation between error rate $E^{(a,b)}$ and $c_{ij}$, which is given by

\begin{equation}
\begin{aligned}
E^{(a,b)}&=\frac{p^{(a,b)}_e+p'^{(a,b)}_e}{Q^{(a,b)}}\\
&=\frac{\big| \tilde{c}_{aa}-\tilde{c}_{bb}\big|^2+\big| \tilde{c}_{ba}-\tilde{c}_{ab}\big|^2+\sum_{i\neq a,b} c^2_{ia}+c^2_{ib}}{2(\sum_i c^2_{ia}+c^2_{ib})}.\\
\end{aligned}
\end{equation}
Furthermore, the error for all sifted key bits is

\begin{equation}
\begin{aligned}
E&=\frac{\sum_{a<b}Q^{(a,b)}E^{(a,b)}}{\sum_{a,b}Q^{(a,b)}}=\frac{\sum_{a<b}\big| \tilde{c}_{aa}-\tilde{c}_{bb}\big|^2+\big| \tilde{c}_{ba}-\tilde{c}_{ab}\big|^2+\sum_{i\neq a,b} c^2_{ia}+c^2_{ib}}{2\sum_{a<b}\sum_i (c^2_{ia}+c^2_{ib})}\\
&\geqslant \frac{\sum_{a<b}\sum_{i\neq a,b} c^2_{ia}+c^2_{ib}}{2\sum_{a<b}\sum_i (c^2_{ia}+c^2_{ib})}=\frac{(L-2)\sum_{i\neq j}c^2_{ij}/2}{(L-1)(\sum_i c^2_{ii}+\sum_{i\neq j}c^2_{ij})}=\frac{(L-2)x_2/2}{(L-1)(x_1+x_2)}.\\
\end{aligned}
\end{equation}
Thus, we have $x_2/(x_1+x_2)\leqslant 2(L-1)E/(L-2)$. In conclusion, with this relation, we can calculate a more tighter bound of $I_{AE}$ with \eqref{Iae1}.

\subsection{Two-photon case}
Alice randomly prepares the two-photon state $\ket{\psi}=\sum^{L}_{i=1}\ket{ii}+\sqrt{2}\sum_{1\leqslant i<j\leqslant L}(-1)^{k_i+k_j}\ket{ij}$, where $k_i,k_j\in\{0,1\}$ is Alice's raw key bit, and $\ket{ij} (i\in\{1,..,L\})$ represents that there is one photon in the $i$-th and $j$-th time-bins respectively. Similar to the single photon case, Eve's general collective attack in two-photon case can be given by:

\begin{equation}
\label{attack2}
\begin{aligned}
&U_{Eve}\ket{ij}\ket{e_{000}}=\sum^L_{l=1}c_{ijl}\ket{l}\ket{e_{ijl}}.\\
\end{aligned}
\end{equation}
When Bob projects the incoming single photon states into $(\ket{a}\pm\ket{b})/\sqrt{2}$ successfully, the evolution of quantum state will be

\begin{equation}
\label{evolution2}
\begin{aligned}
\ket{\psi}\ket{e_{00}}\longrightarrow &\sum_i\tilde{c}_{iia}\ket{a}+\tilde{c}_{iib}\ket{b}+(-1)^{k_a+k_b}\sqrt{2}(\tilde{c}_{aba}\ket{a}+\tilde{c}_{abb}\ket{b})\\
&+\sum_{i\neq a,b}(-1)^{k_i}\sqrt{2}((-1)^{k_a}(\tilde{c}_{iaa}\ket{a}+\tilde{c}_{iab}\ket{b})+(-1)^{k_b}(\tilde{c}_{iba}\ket{a}+\tilde{c}_{ibb}\ket{b}))\\
&+\sum_{i<j,i,j\neq a,b}\sqrt{2}(-1)^{k_i+k_j}(\tilde{c}_{ija}\ket{a}+\tilde{c}_{ijb}\ket{b}).\\
\end{aligned}
\end{equation}
For the ease of presentation, we denote $c_{ijl}\ket{e_{ijl}}$ as $\tilde{c}_{ijl}$, and if $i>j$ for some $c_{ijl}$, we should recognize it as $c_{jil}$. For $\sum_i\tilde{c}_{iil}$, we further simplify it as $\tilde{c}_{l}$.
Clearly, as a result of the random phase $(-1)^{k_i},i\neq a,b$, Eve's state collapses into a mixture state given by
\begin{equation}
\label{evolution22}
\begin{aligned}
\rho^{(a,b)}=&P\{\tilde{c}_a+(-1)^{k_a+k_b}\sqrt{2}\tilde{c}_{aba}\}+P\{\tilde{c}_{b}+(-1)^{k_a+k_b}\sqrt{2}\tilde{c}_{abb}\}\\
&+2\sum_{i\neq a,b} P\{\tilde{c}_{iaa}+(-1)^{k_a+k_b}\tilde{c}_{iba}\}+P\{\tilde{c}_{ibb}+(-1)^{k_a+k_b}\tilde{c}_{iab}\}\\
&+2\sum_{i<j,i,j\neq a,b}P\{\tilde{c}_{ija}\}+P\{\tilde{c}_{ijb}\}.
\end{aligned}
\end{equation}
Now based on very similar considerations in single photon case, we write Eve's information as
\begin{equation}
\label{Iaeab2}
\begin{aligned}
Q^{(a,b)}I^{(a,b)}_{AE}\leqslant&\varphi(\big|\tilde{c}_{a}\big|^2+\big|\tilde{c}_{b}\big|^2,2c^2_{aba}+2c^2_{abb})+\varphi(2\sum_{i\neq a,b}c^2_{iaa}+c^2_{ibb},2\sum_{i\neq a,b}c^2_{iba}+c^2_{iab}).
\end{aligned}
\end{equation}
Furthermore, we have

\begin{equation}
\label{Iae2}
\begin{aligned}
I_{AE}\leqslant&\frac{\sum_{a<b} Q^{(a,b)}I^{(a,b)}_{AE}}{\sum_{a<b}Q^{(a,b)}}\\
&=\frac{\varphi(\sum_{a<b}\big|\tilde{c}_{a}\big|^2+\big|\tilde{c}_{b}\big|^2,\sum_{a<b}2c^2_{aba}+2c^2_{abb})+\varphi(\sum_{a<b}2\sum_{i\neq a,b}c^2_{iaa}+c^2_{ibb},\sum_{a<b}2\sum_{i\neq a,b}c^2_{iba}+c^2_{iab})}{\sum_{a<b}\big|\tilde{c}_{a}\big|^2+\big|\tilde{c}_{b}\big|^2+2\sum_{i<j} c^2_{ija}+c^2_{ijb}}\\&
=\frac{\varphi((L-1)x_1,x_2)+\varphi((L-2)x_2,2x_3)}{(L-1)(x_1+x_2+x_3)},\\
\end{aligned}
\end{equation}
where,
\begin{equation}
\label{defs2}
\begin{aligned}
&x_1\triangleq\sum_i\big|\tilde{c}_{i}\big|^2,\\
&x_2\triangleq\sum_{a<b}2 c^2_{aba}+2c^2_{abb},\\
&x_3\triangleq\sum_{a<b}\sum_{i\neq a,b}2c^2_{abi}.\\
\end{aligned}
\end{equation}

Here, to obtain \eqref{Iae2} we used the Jensen's inequality and the following mathematical observations:
\begin{equation}
\label{relations2}
\begin{aligned}
&\sum_{a<b}\big|\tilde{c}_{a}\big|^2+\big|\tilde{c}_{b}\big|^2=(L-1)x_1,\\
&\sum_{a<b}\sum_{i\neq a,b}2c^2_{iaa}+2c^2_{ibb}=(L-2)x_2,\\
&\sum_{a<b}\sum_{i\neq a,b}2c^2_{iab}+2c^2_{iba}=2x_3,
\end{aligned}
\end{equation}
hold for any non-negative array. Next we try to analyze the restrictions on $x_1$, $x_2$ and $x_3$ with the help of error rate $E$.

We return to Eq.\eqref{evolution2}, it's straightforward to see that the probability for error key events from the first row of Eq.\eqref{evolution2} is

\begin{equation}
\label{error21}
\begin{aligned}
&\frac{1}{2}\sum_{a<b}\big|\tilde{c}_{a}-\tilde{c}_{b}+\sqrt{2}(\tilde{c}_{aba}-\tilde{c}_{abb})\big|^2+\big|\tilde{c}_{a}+\tilde{c}_{b}-\sqrt{2}(\tilde{c}_{aba}+\tilde{c}_{abb})\big|^2\\
&\geqslant \sum_{a<b}(\sqrt{\big|\tilde{c}_{a}\big|^2+\big|\tilde{c}_{b}\big|^2}-\sqrt{2c^2_{aba}+2c^2_{abb}})^2\\
&\geqslant (\sqrt{\sum_{a<b}\big|\tilde{c}_{a}\big|^2+\big|\tilde{c}_{b}\big|^2}-\sqrt{\sum_{a<b}2c^2_{aba}+2c^2_{abb}})^2,\\
&=(\sqrt{(L-1)x_1}-\sqrt{x_2})^2
\end{aligned}
\end{equation}
where, we used the Cauchy-Schwartz inequality twice. And the probability for error-key events from the third row of Eq.\eqref{evolution2} is
\begin{equation}
\label{error22}
\begin{aligned}
&\frac{1}{2}\sum_{a<b}\sum_{i,j\neq a,b}2(c^2_{ija}+c^2_{ijb})=\frac{(L-3)x_3}{2}.
\end{aligned}
\end{equation}
Summing over the Eq.\eqref{error21} and \eqref{error22}, dividing by $\sum_{a<b}Q^{(a,b)}$, we have
\begin{equation}
\label{error2}
\begin{aligned}
E\geqslant \frac{(\sqrt{(L-1)x_1}-\sqrt{x_2})^2+(L-3)x_3/2}{(L-1)(x_1+x_2+x_3)}.
\end{aligned}
\end{equation}
This ends the analyses on two-photon case.

\subsection{Three-photon case}

Alice randomly prepares the three-photon state
\begin{equation}
\label{3photon}
\begin{aligned}
\ket{\psi}=\sum^{L}_{i=1}(-1)^{k_i}(\ket{iii}+\sqrt{3}\sum_{j\neq i}\ket{ijj})+\sqrt{6}\sum_{1\leqslant i<j<l\leqslant L}(-1)^{k_i+k_j+k_l}\ket{ijl},\\
\end{aligned}
\end{equation}
where $k_i,k_j,k_l\in\{0,1\}$ are Alice's raw key bit, and $\ket{ijl} (i,j,l\in\{1,..,L\})$ represents that there is one photon in the $i$-th, $j$-th and $l$-th time-bins respectively. Similar to the single photon case, Eve's general collective attack in three-photon case can be given by:

\begin{equation}
\label{attack2}
\begin{aligned}
&U_{Eve}\ket{ijl}\ket{e_{0000}}=\sum^L_{t=1}c_{ijlt}\ket{t}\ket{e_{ijlt}}.\\
\end{aligned}
\end{equation}
When Bob projects the incoming single photon states into $(\ket{a}\pm\ket{b})/\sqrt{2}$ successfully, the evolution of quantum state will be

\begin{equation}
\label{evolution3}
\begin{aligned}
\ket{\psi}\ket{e_{0000}}\longrightarrow &(-1)^{k_a}(\tilde{c}_{aa}\ket{a}+\tilde{c}_{ab}\ket{b})+(-1)^{k_b}(\tilde{c}_{bb}\ket{b}+\tilde{c}_{ba}\ket{a})\\
&+\sum_{i\neq a,b}(-1)^{k_i}(\tilde{c}_{ia}\ket{a}+\tilde{c}_{ib}\ket{b}+\sqrt{6}(-1)^{k_a+k_b}(\tilde{c}_{iaba}\ket{a}+\tilde{c}_{iabb}\ket{b}))\\
&+\sum_{i,j\neq a,b}(-1)^{k_i+k_j}\sqrt{6}((-1)^{k_a}(\tilde{c}_{ijaa}\ket{a}+\tilde{c}_{ijab}\ket{b})+(-1)^{k_b}(\tilde{c}_{ijba}\ket{a}+\tilde{c}_{ijbb}\ket{b}))\\
&+\sum_{i,j,l\neq a,b}(-1)^{k_i+k_j+k_l}\sqrt{6}(\tilde{c}_{ijla}\ket{a}+\tilde{c}_{ijlb}\ket{b}),
\end{aligned}
\end{equation}
where, $\tilde{c}_{ij}\triangleq \tilde{c}_{iiij}+\sqrt{3}\sum_{t\neq i}\tilde{c}_{ittj}$, and $\tilde{c}_{ijlt}\triangleq c_{ijlt}\ket{e_{ijlt}}$. We have observed that its first row and third row have very similar same form with the evolution of single photon given by Eq.\eqref{evolution1}, while the second row has the similar form with the first row of Eq.\eqref{evolution2}. Thus, analogous to the calculations in single photon and two-photon cases, we have

\begin{equation}
\label{Iae3}
\begin{aligned}
I_{AE}\leqslant&\frac{\sum_{a<b} Q^{(a,b)}I^{(a,b)}_{AE}}{\sum_{a<b}Q^{(a,b)}}=\frac{\varphi((L-1)x_1,x_2)+\varphi((L-2)x_2,2x_3)+\varphi((L-3)x_3,3x_4)}{(L-1)(x_1+x_2+x_3+x_4)},\\
\end{aligned}
\end{equation}
where,
\begin{equation}
\label{defs3}
\begin{aligned}
&x_1\triangleq\sum_i\big|\tilde{c}_{ii}\big|^2,\\
&x_2\triangleq\sum_{i\neq j}\big|\tilde{c}_{ij}\big|^2,\\
&x_3\triangleq\sum_{i<j<l}6c^2_{ijli}+6c^2_{ijlj}+6c^2_{ijll},\\
&x_4\triangleq\sum_{i<j<l}\sum_{t\neq i,j,l}6c^2_{ijlt}.\\
\end{aligned}
\end{equation}

Here, to obtain \eqref{Iae3} we used the Jensen's inequality and the following mathematical observations:
\begin{equation}
\label{relations3}
\begin{aligned}
&\sum_{a<b}\big|\tilde{c}_{aa}\big|^2+\big|\tilde{c}_{bb}\big|^2=(L-1)x_1,\\
&\sum_{a<b}\sum_{i\neq a,b}\big|\tilde{c}_{ia}\big|^2+\big|\tilde{c}_{ib}\big|^2=(L-2)x_2,\\
&\sum_{a<b}\sum_{i,j\neq a,b}6c^2_{ijaa}+6c^2_{ijbb}=(L-3)x_3,\\
&\sum_{a<b}\sum_{i\neq a,b}6 c^2_{iaba}+6c^2_{iabb}=2x_3\\
&\sum_{a<b}\sum_{i,j\neq a,b}6c^2_{ijab}+6c^2_{ijba}=3x_4,
\end{aligned}
\end{equation}
hold for any non-negative array. Next we try to analyze the restrictions on $x_1$, $x_2$ and $x_3$ with the help of error rate $E$.
Based on similar method in last subsection, we have
\begin{equation}
\label{error3}
\begin{aligned}
E\geqslant \frac{(\sqrt{(L-2)x_2}-\sqrt{2x_3})^2+(L-4)x_4/2}{(L-1)(x_1+x_2+x_3+x_4)},
\end{aligned}
\end{equation}
where, we used Cauchy-Schwartz inequality and the following mathematical identity
\begin{equation}
\label{relations32}
\begin{aligned}
&\sum_{a<b}\sum_{i<j<l,i,j,l\neq a,b}6c^2_{ijla}+6c^2_{ijlb}=(L-4)x_4,
\end{aligned}
\end{equation}
always holds.

This ends the analyses on two-photon case.
\subsection{Four-photon case}
Alice randomly prepares the four-photon state
\begin{equation}
\label{3photon}
\begin{aligned}
\ket{\psi}=\sum^{L}_{i=1}\ket{iiii}+\sum_{i<j}\ket{iijj}+(-1)^{k_i+k_j}(\ket{ijjj}+\ket{iiij}+\sum_{n\neq i,j}\ket{ijnn})+\sum_{i<j<l<m}(-1)^{k_i+k_j+k_l+k_m}\ket{ijlm},\\
\end{aligned}
\end{equation}
where we treat the efficiencies as part of quantum state, e.g., $a\ket{ijlm}$ is simply denoted by $\ket{ijlm}$.
Similar to the single photon case, Eve's general collective attack in four-photon case can be given by:
\begin{equation}
\label{attack2}
\begin{aligned}
&U_{Eve}\ket{ijlm}\ket{e_{00000}}=\sum^L_{t=1}c_{ijlmt}\ket{t}\ket{e_{ijlmt}}.\\
\end{aligned}
\end{equation}
When Bob projects the incoming single photon states into $(\ket{a}\pm\ket{b})/\sqrt{2}$ successfully, the evolution of quantum state will be
\begin{equation}
\label{evolution4}
\begin{aligned}
\ket{\psi}\ket{e_{00000}}\longrightarrow &(\tilde{c}_{a}\ket{a}+\tilde{c}_{b}\ket{b})+(-1)^{k_a+k_b}(\tilde{c}_{aba}\ket{a}+\tilde{c}_{abb}\ket{b})\\
&+\sum_{i\neq a,b}(-1)^{k_i}(\tilde{c}_{iaa}\ket{a}+\tilde{c}_{iab}\ket{b}+(-1)^{k_a+k_b}(\tilde{c}_{iba}\ket{a}+\tilde{c}_{ibb}\ket{b}))\\
&+\sum_{i,j\neq a,b}(-1)^{k_i+k_j}((\tilde{c}_{ija}\ket{a}+\tilde{c}_{ijb}\ket{b})+(-1)^{k_a+k_b}(\tilde{c}_{ijaba}\ket{a}+\tilde{c}_{ijabb}\ket{b}))\\
&+\sum_{i,j,l\neq a,b}(-1)^{k_i+k_j+k_l}(\tilde{c}_{ijlaa}\ket{a}+\tilde{c}_{ijlab}\ket{b}+(-1)^{k_a+k_b}(\tilde{c}_{ijlba}\ket{a}+\tilde{c}_{ijlbb}\ket{b}))\\
&+\sum_{i,j,l,m\neq a,b}(-1)^{k_i+k_j+k_l+k_m}(\tilde{c}_{ijlma}\ket{a}+\tilde{c}_{ijlmb}\ket{b}),
\end{aligned}
\end{equation}
where, $\tilde{c}_t\triangleq\sum^{L}_{i=1}\tilde{c}_{iiiit}+\sum_{i<j}\tilde{c}_{iijjt}$,$\tilde{c}_{ijt}\triangleq \tilde{c}_{ijjjt}+\tilde{c}_{iiijt}+\sum_{n\neq i,j}\tilde{c}_{ijnnt}$,and $\tilde{c}_{ijlmt}=c_{ijlmt}\ket{e_{ijlmt}}$. Clearly, each row has the same pattern with evolution of two-photon case. Based on similar techniques used in last three subsections, we have
\begin{equation}
\label{Iae4}
\begin{aligned}
I_{AE}\leqslant&=\frac{\varphi((L-1)x_1,x_2)+\varphi((L-2)x_2,2x_3)+\varphi((L-3)x_3,3x_4)+\varphi((L-4)x_4,4x_5)}{(L-1)(x_1+x_2+x_3+x_4+x_5)},\\
\end{aligned}
\end{equation}
where,
\begin{equation}
\label{defs3}
\begin{aligned}
&x_1\triangleq\sum_i\big|\tilde{c}_{i}\big|^2,\\
&x_2\triangleq\sum_{a<b}\big|\tilde{c}_{aba}\big|^2+\big|\tilde{c}_{abb}\big|^2,\\
&x_3\triangleq\sum_{a < b}\sum_{i\neq a,b}\big|\tilde{c}_{abi}\big|^2,\\
&x_4\triangleq\sum_{i<j<l<m}c^2_{ijlmi}+c^2_{ijlmj}+c^2_{ijlml}+c^2_{ijlmm},\\
&x_5\triangleq\sum_{i<j<l<m}\sum_{t\neq i,j,l,m}c^2_{ijlmt}.\\
\end{aligned}
\end{equation}
And these parameters are constrained by the error rate $E$,
\begin{equation}
\label{error4}
\begin{aligned}
E\geqslant \frac{(\sqrt{(L-1)x_1}-\sqrt{x_2})^2+(\sqrt{(L-3)x_3}-\sqrt{3x_4})^2+(L-5)x_5/2}{(L-1)(x_1+x_2+x_3+x_4+x_5)}.
\end{aligned}
\end{equation}
This ends the analyses on four-photon case.

\subsection{Odd photon-number case}
Alice randomly prepares an encoding state like before, but the photon-number $N$ is an odd number and $L\geqslant N+1$. It is clear that her encoding state has the form
\begin{equation}
\label{oddphoton}
\begin{aligned}
\ket{\psi}=&\sum_{i_1}(-1)^{k_{i_1}}\ket{i_1}+\sum_{i_1<i_2<i_3}(-1)^{k_{i_1}+k_{i_2}+k_{i_3}}\ket{i_1i_2i_3}+\sum_{i_1<i_2<i_3<i_4<i_5}(-1)^{k_{i_1}+k_{i_2}+k_{i_3}+k_{i_4}+k_{i_5}}\ket{i_1i_2i_3i_4i_5}\\
&+...+\sum_{i_1<i_2<i_3<...<N}(-1)^{k_{i_1}+k_{i_2}+k_{i_3}+...+k_{i_N}}\ket{i_1i_2i_3...i_N},
\end{aligned}
\end{equation}
i.e., it consists of n-phase ($n=1,3,5,...,N$) state, denoted by $\ket{i_1i_2..i_n}$. For example, $\ket{i_1i_2...i_n}$ represents that the photon number in time-bins $i_1$, $i_2$,..., and $i_n$ must be odd, while the photon numbers in all other time-bins must be even.
Eve's general collective attack in this case can be given by:
\begin{equation}
\label{attackodd}
\begin{aligned}
&U_{Eve}\ket{i_1i_2..i_n}\ket{e_{initial}}=\sum^L_{t=1}c_{i_1i_2..i_nt}\ket{t}\ket{e_{i_1i_2..i_nt}}\triangleq \sum^L_{t=1}\tilde{c}_{i_1i_2..i_nt}\ket{t}.\\
\end{aligned}
\end{equation}
When Bob projects the incoming single photon states into $(\ket{a}\pm\ket{b})/\sqrt{2}$ successfully, the evolution of quantum state will be
\begin{equation}
\label{evolutionodd}
\begin{aligned}
\ket{\psi}\ket{e_{initial}}\longrightarrow &
(-1)^{k_a}(\tilde{c}_{aa}+\tilde{c}_{ab})+(-1)^{k_b}(\tilde{c}_{ba}+\tilde{c}_{bb})\\
&+\sum_{i_1\neq a,b}(-1)^{i_1}(\tilde{c}_{i_1a}\ket{a}+\tilde{c}_{i_1b}\ket{b})+(-1)^{k_a+k_b}(\tilde{c}_{i_1aba}\ket{a}+\tilde{c}_{i_1abb}\ket{b})\\
&+\sum_{i_1i_2\neq a,b}(-1)^{k_{i_1}+k_{i_2}}((\tilde{c}_{i_1i_2aa}\ket{a}+\tilde{c}_{i_1i_2ab}\ket{b})+(-1)^{k_a+k_b}(\tilde{c}_{i_1i_2ba}\ket{a}+\tilde{c}_{i_1i_2bb}\ket{b}))\\
&+...\\
&
\begin{aligned}
+\sum_{i_1...i_{N-2}\neq a,b}(-1)^{k_{i_1}+...+k_{i_{N-2}}}((&\tilde{c}_{i_1...i_{N-2}a}\ket{a}+\tilde{c}_{i_1...i_{N-2}b}\ket{b})\\
&+(-1)^{k_a+k_b}(\tilde{c}_{i_1...i_{N-2}aba}\ket{a}+\tilde{c}_{i_1...i_{N-2}abb}\ket{b}))\\
\end{aligned}
\\
&
\begin{aligned}
+\sum_{i_1...i_{N-1}\neq a,b}(-1)^{k_{i_1}+k_{i_2}+..+k_{i_{N-1}}}((&\tilde{c}_{i_1...i_{N-1}aa}\ket{a}+\tilde{c}_{i_1...i_{N-1}ab}\ket{b})\\
&+(-1)^{k_a+k_b}(\tilde{c}_{i_1...i_{N-1}ba}\ket{a}+\tilde{c}_{i_1...i_{N-1}bb}\ket{b}))
\end{aligned}
\\
&+\sum_{i_1...i_N\neq a,b}(-1)^{k_{i_1}+...+k_{i_N}}(\tilde{c}_{i_1...i_Na}\ket{a}+\tilde{c}_{i_1...i_Nb}\ket{b}).
\end{aligned}
\end{equation}
Evidently, for each summation we can calculate Eve's information. Specifically, for the summations with the global phase $(-1)^{k_{i_1}+..+k_{i_n}}$ and $n$ is odd, we obtain
\begin{equation}
\label{Iaeoddodd}
\begin{aligned}
Q^{(a,b)}I^{(a,b)}_{AE}\leqslant&\sum_{i_1...i_n\neq a,b}\varphi(\big|\tilde{c}_{i_1...i_na}\big|^2+\big|\tilde{c}_{i_1...i_nb}\big|^2,\big|\tilde{c}_{i_1...i_naba}\big|^2+\big|\tilde{c}^2_{i_1...+i_nabb}\big|^2).
\end{aligned}
\end{equation}
For the summations with the global phase $(-1)^{k_{i_1}+..+k_{i_n}}$ and $n$ is even, we obtain
\begin{equation}
\label{Iaeoddeven}
\begin{aligned}
Q^{(a,b)}I^{(a,b)}_{AE}\leqslant&\sum_{i_1...i_n\neq a,b}\varphi(\big|\tilde{c}_{i_1...i_naa}\big|^2+\big|\tilde{c}_{i_1...i_nbb}\big|^2,\big|\tilde{c}_{i_1...i_nab}\big|^2+\big|\tilde{c}^2_{i_1...+i_nba}\big|^2).
\end{aligned}
\end{equation}
Noting the following mathematical identities
\begin{equation}
\label{relationsodd}
\begin{aligned}
&\sum_{a<b}\big|\tilde{c}_{aa}\big|^2+\big|\tilde{c}_{bb}\big|^2=(L-1)\sum_{i}\big|\tilde{c}_{ii}\big|^2,\\
&\sum_{a<b}\big|\tilde{c}_{ab}\big|^2+\big|\tilde{c}_{ba}\big|^2=\sum_{i\neq j} \big|\tilde{c}_{ij}\big|^2,\\
&\sum_{a<b}\sum_{i_1...i_n\neq a,b}\big|\tilde{c}_{i_1...i_na}\big|^2+\big|\tilde{c}_{i_1...i_nb}\big|^2=(L-n-1)\sum_{i_1...i_n}\sum_{t\neq i_1...i_n} \big|\tilde{c}_{i_1...i_nt}\big|^2,\\
&\sum_{a<b}\sum_{i_1...i_n\neq a,b}\big|\tilde{c}_{i_1...i_naa}\big|^2+\big|\tilde{c}_{i_1...i_nbb}\big|^2=(L-n-1)\sum_{i_1...i_{n+1}}\sum_{t= i_1}^{i_{n+1}} c_{i_1...i_{n+1}t},\\
&\sum_{a<b}\sum_{i_1...i_{n+1}\neq a,b}\big|\tilde{c}_{i_1...i_nba}\big|^2+\big|\tilde{c}_{i_1...i_nab}\big|^2=(n+1)\sum_{i_1...i_{n+1}}\sum_{t\neq i_1...i_{n+1}}c_{i_1...i_{n+1}t},\\
&\sum_{a<b}\sum_{i_1...i_n\neq a,b}\big|\tilde{c}_{i_1...i_naba}\big|^2+\big|\tilde{c}_{i_1...i_nabb}\big|^2=(n+1)\sum_{i_1...i_{n+2}}\sum_{t= i_1}^{i_{n+2}} c_{i_1...i_{n+2}t}.\\
\end{aligned}
\end{equation}
And define
\begin{equation}
\label{defsodd}
\begin{aligned}
&x_1\triangleq \sum_i \big|\tilde{c}_{ii}\big|^2,\\
&x_2\triangleq \sum_{i\neq j} \big|\tilde{c}_{ij}\big|^2,\\
& x_{n} \triangleq \sum_{i_1...i_n}\sum_{t= i_1}^{i_n} \big|\tilde{c}_{i_1...i_nt}\big|^2,\\
& x_{n+1} \triangleq \sum_{i_1...i_n}\sum_{t\neq i_1...i_n} \big|\tilde{c}_{i_1...i_nt}\big|^2.\\
\end{aligned}
\end{equation}
Combining Eqs.\eqref{Iaeoddodd}, \eqref{Iaeoddeven}, \eqref{relationsodd} and \eqref{defsodd}, we have
\begin{equation}
\label{Iaeodd1}
\begin{aligned}
\sum_{a<b}Q^{(a,b)}I^{(a,b)}_{AE}\leqslant& \varphi((L-1)x_1,x_2)+\varphi((L-2)x_2,2x_3)+\varphi((L-3)x_3,3x_4)+...+\varphi((L-N)x_N,Nx_{N+1}).
\end{aligned}
\end{equation}
Besides, with Eqs.\eqref{oddphoton}, \eqref{attackodd} and \eqref{defsodd}, it's easy to verify $\sum_{a<b}Q^{(a,b)}=(L-1)(x_1+x_2+...+x_{N+1})$.
In conclusion, Eve's information is
\begin{equation}
\label{Iaeodd}
\begin{aligned}
I_{AE}=\frac{\sum_{a<b}Q^{(a,b)}I^{(a,b)}_{AE}}{\sum_{a<b}Q^{(a,b)}}\leqslant& \frac{\sum_{n=1}^{N}\varphi((L-n)x_n,nx_{n+1})}{(L-1)\sum_{n=1}^{N+1}x_n}.
\end{aligned}
\end{equation}
Through calculating the probabilities of error-key events corresponds to the $n$-th ($n$ is even) row of Eq.\eqref{evolutionodd}, we obtain that the error rate $E$ must satisfy
\begin{equation}
\label{errorodd}
\begin{aligned}
E\geqslant \frac{\sum^{(N-1)/2}_{n\geqslant 1}(\sqrt{(L-2n)x_{2n}}-\sqrt{2nx_{2n+1}})^2+(L-N-1)x_{N+1}/2}{(L-1)\sum_{n=1}^{N+1}x_n}.
\end{aligned}
\end{equation}
This ends the security proof for odd photon-number case.
\subsection{Even photon-number case}
Alice randomly prepares an encoding state like before, but the photon-number $N$ is an even number and $L\geqslant N+1$. It is clear that her encoding state has the form
\begin{equation}
\label{evenphoton}
\begin{aligned}
\ket{\psi}=&\ket{i_0}+\sum_{i_1<i_2}(-1)^{k_{i_1}+k_{i_2}}\ket{i_1i_2}+\sum_{i_1<i_2<i_3<i_4}(-1)^{k_{i_1}+k_{i_2}+k_{i_3}+k_{i_4}}\ket{i_1i_2i_3i_4}\\
&+...+\sum_{i_1<i_2<i_3<...<i_N}(-1)^{k_{i_1}+k_{i_2}+k_{i_3}+...+k_{i_N}}\ket{i_1i_2i_3...i_N},
\end{aligned}
\end{equation}
i.e., it consists of n-phase ($n=0,2,4,...,N$) state, denoted by $\ket{i_1i_2..i_n}$. For example, $\ket{i_1i_2...i_n}$ represents that the photon number in time-bins $i_1$, $i_2$,..., and $i_n$ must be odd, while the photon numbers in all other time-bins must be even.
Eve's general collective attack in this case can be given by:
\begin{equation}
\label{attackeven}
\begin{aligned}
&U_{Eve}\ket{i_1i_2..i_n}\ket{e_{initial}}=\sum^L_{t=1}c_{i_1i_2..i_nt}\ket{t}\ket{e_{i_1i_2..i_nt}}\triangleq \sum^L_{t=1}\tilde{c}_{i_1i_2..i_nt}\ket{t}.\\
\end{aligned}
\end{equation}
When Bob projects the incoming single photon states into $(\ket{a}\pm\ket{b})/\sqrt{2}$ successfully, the evolution of quantum state will be
\begin{equation}
\label{evolutioneven}
\begin{aligned}
\ket{\psi}\ket{e_{initial}}\longrightarrow &
(\tilde{c}_{i_0a}\ket{a}+\tilde{c}_{i_0b}\ket{b})+(-1)^{k_a+k_b}(\tilde{c}_{aba}\ket{a}+\tilde{c}_{abb}\ket{b})\\
&+\sum_{i_1\neq a,b}(-1)^{i_1}(\tilde{c}_{i_1aa}\ket{a}+\tilde{c}_{i_1ab}\ket{b})+(-1)^{k_a+k_b}(\tilde{c}_{i_1ba}\ket{a}+\tilde{c}_{i_1bb}\ket{b})\\
&+\sum_{i_1i_2\neq a,b}(-1)^{k_{i_1}+k_{i_2}}((\tilde{c}_{i_1i_2a}\ket{a}+\tilde{c}_{i_1i_2b}\ket{b})+(-1)^{k_a+k_b}(\tilde{c}_{i_1i_2aba}\ket{a}+\tilde{c}_{i_1i_2abb}\ket{b}))\\
&+...\\
&
\begin{aligned}
+\sum_{i_1...i_{N-2}\neq a,b}(-1)^{k_{i_1}+...+k_{i_{N-2}}}((&\tilde{c}_{i_1...i_{N-2}a}\ket{a}+\tilde{c}_{i_1...i_{N-2}b}\ket{b})\\
&+(-1)^{k_a+k_b}(\tilde{c}_{i_1...i_{N-2}aba}\ket{a}+\tilde{c}_{i_1...i_{N-2}abb}\ket{b}))\\
\end{aligned}
\\
&
\begin{aligned}
+\sum_{i_1...i_{N-1}\neq a,b}(-1)^{k_{i_1}+k_{i_2}+..+k_{i_{N-1}}}((&\tilde{c}_{i_1...i_{N-1}aa}\ket{a}+\tilde{c}_{i_1...i_{N-1}ab}\ket{b})\\
&+(-1)^{k_a+k_b}(\tilde{c}_{i_1...i_{N-1}ba}\ket{a}+\tilde{c}_{i_1...i_{N-1}bb}\ket{b}))
\end{aligned}
\\
&+\sum_{i_1...i_N\neq a,b}(-1)^{k_{i_1}+...+k_{i_N}}(\tilde{c}_{i_1...i_Na}\ket{a}+\tilde{c}_{i_1...i_Nb}\ket{b}).
\end{aligned}
\end{equation}
Evidently, for each summation we can calculate Eve's information. Specifically, for the summations with the global phase $(-1)^{k_{i_1}+..+k_{i_n}}$ and $n$ is even, we obtain
\begin{equation}
\label{Iaeeveneven}
\begin{aligned}
Q^{(a,b)}I^{(a,b)}_{AE}\leqslant&\varphi(\big|\tilde{c}_{i_0a}\big|^2+\big|\tilde{c}_{i_0b}\big|^2,\big|\tilde{c}_{aba}\big|^2+\big|\tilde{c}^2_{abb}\big|^2)\\
&\sum_{i_1...i_n\neq a,b}\varphi(\big|\tilde{c}_{i_1...i_na}\big|^2+\big|\tilde{c}_{i_1...i_nb}\big|^2,\big|\tilde{c}_{i_1...i_naba}\big|^2+\big|\tilde{c}^2_{i_1...+i_nabb}\big|^2).
\end{aligned}
\end{equation}
For the summations with the global phase $(-1)^{k_{i_1}+..+k_{i_n}}$ and $n$ is odd, we obtain
\begin{equation}
\label{Iaeevenodd}
\begin{aligned}
Q^{(a,b)}I^{(a,b)}_{AE}\leqslant&\sum_{i_1...i_n\neq a,b}\varphi(\big|\tilde{c}_{i_1...i_naa}\big|^2+\big|\tilde{c}_{i_1...i_nbb}\big|^2,\big|\tilde{c}_{i_1...i_nab}\big|^2+\big|\tilde{c}^2_{i_1...+i_nba}\big|^2).
\end{aligned}
\end{equation}
Noting the following mathematical identities
\begin{equation}
\label{relationseven}
\begin{aligned}
&\sum_{a<b}\big|\tilde{c}_{i_0a}\big|^2+\big|\tilde{c}_{i_0b}\big|^2=(L-1)\sum_{j}\big|\tilde{c}_{i_0j}\big|^2,\\
&\sum_{a<b}\big|\tilde{c}_{aba}\big|^2+\big|\tilde{c}_{abb}\big|^2=\sum_{i_1 < i_2} \big|\tilde{c}_{i_1i_2i_1}\big|^2+\big|\tilde{c}_{i_1i_2i_2}\big|^2,\\
&\sum_{a<b}\sum_{i_1...i_n\neq a,b}\big|\tilde{c}_{i_1...i_na}\big|^2+\big|\tilde{c}_{i_1...i_nb}\big|^2=(L-n-1)\sum_{i_1...i_n}\sum_{t\neq i_1...i_n} \big|\tilde{c}_{i_1...i_nt}\big|^2,\\
&\sum_{a<b}\sum_{i_1...i_n\neq a,b}\big|\tilde{c}_{i_1...i_naa}\big|^2+\big|\tilde{c}_{i_1...i_nbb}\big|^2=(L-n-1)\sum_{i_1...i_{n+1}}\sum_{t= i_1}^{i_{n+1}} c_{i_1...i_{n+1}t},\\
&\sum_{a<b}\sum_{i_1...i_{n+1}\neq a,b}\big|\tilde{c}_{i_1...i_nba}\big|^2+\big|\tilde{c}_{i_1...i_nab}\big|^2=(n+1)\sum_{i_1...i_{n+1}}\sum_{t\neq i_1...i_{n+1}}c_{i_1...i_{n+1}t},\\
&\sum_{a<b}\sum_{i_1...i_n\neq a,b}\big|\tilde{c}_{i_1...i_naba}\big|^2+\big|\tilde{c}_{i_1...i_nabb}\big|^2=(n+1)\sum_{i_1...i_{n+2}}\sum_{t= i_1}^{i_{n+2}} c_{i_1...i_{n+2}t}.\\
\end{aligned}
\end{equation}
And define
\begin{equation}
\label{defseven}
\begin{aligned}
&x_1\triangleq \sum_{j}\big|\tilde{c}_{i_0j}\big|^2,\\
&x_2\triangleq \sum_{i < j} \big|\tilde{c}_{iji}\big|^2+\big|\tilde{c}_{ijj}\big|^2,\\
& x_{n} \triangleq \sum_{i_1...i_n}\sum_{t= i_1}^{i_n} \big|\tilde{c}_{i_1...i_nt}\big|^2,\\
& x_{n+1} \triangleq \sum_{i_1...i_n}\sum_{t\neq i_1...i_n} \big|\tilde{c}_{i_1...i_nt}\big|^2.\\
\end{aligned}
\end{equation}
Combining Eqs.\eqref{Iaeeveneven}, \eqref{Iaeevenodd}, \eqref{relationseven} and \eqref{defseven}, we have
\begin{equation}
\label{Iaeeven1}
\begin{aligned}
\sum_{a<b}Q^{(a,b)}I^{(a,b)}_{AE}\leqslant& \varphi((L-1)x_1,x_2)+\varphi((L-2)x_2,2x_3)+\varphi((L-3)x_3,3x_4)+...+\varphi((L-N)x_N,Nx_{N+1}).
\end{aligned}
\end{equation}
Besides, with Eqs.\eqref{evenphoton}, \eqref{attackeven} and \eqref{defseven}, it's easy to verify $\sum_{a<b}Q^{(a,b)}=(L-1)(x_1+x_2+...+x_{N+1})$.
In conclusion, Eve's information is
\begin{equation}
\label{Iaeeven}
\begin{aligned}
I_{AE}=\frac{\sum_{a<b}Q^{(a,b)}I^{(a,b)}_{AE}}{\sum_{a<b}Q^{(a,b)}}\leqslant& \frac{\sum_{n=1}^{N}\varphi((L-n)x_n,nx_{n+1})}{(L-1)\sum_{n=1}^{N+1}x_n}.
\end{aligned}
\end{equation}
Through calculating the probabilities of error-key events corresponds to the $n$-th ($n$ is odd) row of Eq.\eqref{evolutioneven}, we obtain that the error rate $E$ must satisfy
\begin{equation}
\label{errorodd}
\begin{aligned}
E\geqslant \frac{\sum^{N/2}_{n\geqslant 1}(\sqrt{(L-2n+1)x_{2n-1}}-\sqrt{(2n-1)x_{2n}})^2+(L-N-1)x_{N+1}/2}{(L-1)\sum_{n=1}^{N+1}x_n}.
\end{aligned}
\end{equation}
This ends the security proof for even photon-number case.

\subsection{General $N$-photon case}
We summarize and simplify the results given by the even photon-number and odd photon-number cases here. For a RRDPS protocol with $N$ photon-number source, packet size $L$ and $L\geqslant N+1$, Eve's information can by bounded by
\begin{equation}
\label{Iae}
\begin{aligned}
I_{AE}\leqslant& \frac{\sum_{n=1}^{N}\varphi((L-n)x_n,nx_{n+1})}{L-1},
\end{aligned}
\end{equation}
where, $\varphi(x,y)=-x\log_2{x}-y\log_2{y}+(x+y)\log_2{(x+y)}$, and non-negative real parameters $x_i$ satisfying $\sum_{i=1}^{N+1}x_i=1$. If Alice and Bob make sure that their error rate is $E$, then the parameters will also satisfy that:
\begin{equation}
\label{erroroddcons}
\begin{aligned}
&\text{if $N$ is odd},\\
&E\geqslant \frac{\sum^{(N-1)/2}_{n\geqslant 1}(\sqrt{(L-2n)x_{2n}}-\sqrt{2nx_{2n+1}})^2+(L-N-1)x_{N+1}/2}{L-1};\\
&\text{if $N$ is even},\\
&E\geqslant \frac{\sum^{N/2}_{n\geqslant 1}(\sqrt{(L-2n+1)x_{2n-1}}-\sqrt{(2n-1)x_{2n}})^2+(L-N-1)x_{N+1}/2}{L-1}.\\
\end{aligned}
\end{equation}

Base one above results, a corollary is straightforward which is: for any $N<L-1$, $I_{AE}<1$ holds. Lets prove this corollary by reduction to absurdity. We consider $I_{AE}=1$ in case of $N<L-1$. According to the property of $\varphi(x,y)$ function,

\begin{equation}
\label{corollary1}
\begin{aligned}
1=I_{AE}\leqslant& \frac{\sum_{n=1}^{N}\varphi((L-n)x_n,nx_{n+1})}{L-1}\leqslant \frac{\sum_{n=1}^{N}(L-n)x_n+nx_{n+1}}{L-1}=\frac{\sum_{n=1}^{N}(L-1)x_n+Nx_{N+1}}{L-1}\\
&=1-x_{N+1}+\frac{N}{L-1}x_{N+1}=1+\frac{N-(L-1)}{L-1}x_{N+1}.
\end{aligned}
\end{equation}
Evidently, this suggests that $x_{N+1}=0$, which leads to $\varphi((L-N)x_N, x_{N+1})=0$. Then Eq.\eqref{corollary1} is rewritten as
\begin{equation}
\label{corollary2}
\begin{aligned}
1=I_{AE}\leqslant& \frac{\sum_{n=1}^{N-1}\varphi((L-n)x_n,nx_{n+1})}{L-1}\leqslant \frac{\sum_{n=1}^{N-1}(L-1)x_n+Nx_{N}}{L-1}\\
&=1-x_{N}+\frac{N-1}{L-1}x_{N}=1+\frac{N-1-(L-1)}{L-1}x_{N},
\end{aligned}
\end{equation}
which implies $x_N=0$. Repeat above arguments for $N$ times, we obtain that $x_n=0 (n=2,3,4..)$ and $I_{AE}=0$, which conflicts with $I_{AE}=1$. This ends the proof of this corollary.

\end{document}